\documentclass[12pt]{article}
\usepackage{latexsym}
\usepackage{graphicx}
\usepackage{caption}
\usepackage{psfrag}
\usepackage{amsmath,amssymb,dsfont}
\newcommand{\be}{\begin{equation}}
\newcommand{\ee}{\end{equation}}
\newcommand{\bea}{\begin{eqnarray}}
\newcommand{\eea}{\end{eqnarray}}

\newcommand{\nn}{\nonumber}

\input epsf

\numberwithin{equation}{section}

\begin{document}

\begin{titlepage}

\begin{flushright}
\end{flushright}
\vspace*{1.5cm}
\begin{center}
{\Large \bf Possible duality violations in $\tau$ decay and their impact on the  determination of $\alpha_s$ }\\[3.0cm]

{\bf Oscar Cat\`a,$^{1}$  Maarten Golterman,$^{2}$  and Santiago Peris$^{3}$} \\[1cm]
$^{1}$INFN, Laboratori Nazionali di Frascati, Via E. Fermi 40, I-00044 Frascati, Italy\\
$^{2}$Department of Physics and Astronomy, San Francisco State
University,\\ 1600 Holloway Ave, San Francisco, CA 94132, USA\\
$^{3}$Grup de F{\'\i}sica Te{\`o}rica and IFAE, UAB, E-08193 Bellaterra, Barcelona, Spain\\[0.5cm]

\end{center}

\vspace*{1.0cm}

\begin{abstract}

We discuss the issue of duality violations in hadronic tau decay. After introducing a physically motivated {\it ansatz} for duality violations, we estimate their possible size by fitting this {\it ansatz} to the tau experimental data provided by the ALEPH collaboration. Our conclusion is that these data do not exclude significant duality violations in tau decay. This may imply an additional systematic error in the value of $\alpha_s(m_{\tau})$, extracted from tau decay,
as large as $\delta\alpha_s(m_{\tau})\sim 0.003-0.010$ .

\end{abstract}

\end{titlepage}

\setcounter{footnote}{0}

\section{Introduction}

The hadronic decay of the tau lepton is a particularly suitable process for studying QCD interactions, because
its mass is heavy enough to produce hadrons in the final state, without the
complications which arise with hadrons in the initial state.
However, since the mass of the tau  is
not much larger than a typical hadronic scale of order $1$~GeV,
it was initially unclear whether a perturbative QCD analysis of this process
would not be ruined by
nonperturbative effects beyond any systematic control. The pioneering work of
Ref.~\cite{BNP} showed, however, that perturbation theory, augmented with the
Operator
Product Expansion (OPE)~\cite{SVZ}, can indeed be the right tool for
understanding strong interaction aspects of hadronic tau decay.
Among these, the most important ones are perhaps the determination of the coupling constant $\alpha_s$, the light quark masses, and the vacuum condensates~\cite{the works}.

In fact, the determination of $\alpha_s$ from tau decay is now one of the most accurate ones. The level of accuracy is so high that subtle effects, which were completely negligible before, may now  start to play a significant role in the
precision of the result. Since the value of $\alpha_s$ is of fundamental
importance for our understanding of QCD and the Standard Model, it is obvious
that a good control of all systematic effects is, now more than ever, mandatory.

References~\cite{Davier, Baikov, Maltman, Jamin} find the following values for $\alpha_s$ after analyzing the same experimental data obtained by ALEPH~\cite{Aleph}:
\begin{eqnarray}\label{alphas}
\alpha_s(m_{\tau}^2)&=&0.344\pm 0.005_{\mathrm{exp}}\pm 0.007_{\mathrm{th}}\quad \cite{Davier} \ , \nn \\
\alpha_s(m_{\tau}^2)&=&0.332\pm 0.005_{\mathrm{exp}}\pm 0.015_{\mathrm{th}}\quad \cite{Baikov} \ , \nn \\
\alpha_s(m_{\tau}^2)&=&0.321\pm 0.005_{\mathrm{exp}}\pm 0.012_{\mathrm{th}}\quad \cite{Maltman}\ , \nn \\
\alpha_s(m_{\tau}^2)&=&0.316\pm 0.003_{\mathrm{exp}}\pm 0.005_{\mathrm{th}}\quad \cite{Jamin}\ .
\end{eqnarray}
Note that, if the errors are taken at face value, these four results
are not fully consistent with one another. The main difference between the results
of Refs.~\cite{Davier} and \cite{Jamin} is due to the treatment of
perturbation theory (the use of so-called Contour-Improved (CI) {\it vs} Fixed-Order (FO)
resummation prescriptions, respectively \cite{Pich}). This difference has been included in the systematic error in \cite{Baikov}. The difference between the
result of Ref.~\cite{Maltman} and the other three is mainly due to the use of different
weight functions in the finite-energy sum rules. Different choices for these
functions distribute differently the relative weight between the perturbative
and the OPE contributions, as extracted from the
spectral function. Although in an exact treatment these choices should not
matter, in practice they do since truncations of both the perturbative
series and higher orders in the OPE have to be  applied. Ref.~\cite{Maltman} also analyzes the effect of some of these higher orders in the OPE.

Another interesting ``mismatch'' is seen in the analysis of the
ALEPH data carried out in Ref.~\cite{Davier}, which quotes the following values
for the  gluon condensate:
\begin{eqnarray} \label{gluonVA}
\frac{\alpha_s}{\pi}\langle GG\rangle\Big|_{\mathrm{Vector}}\!\!&=&(-0.8\pm0.4)\times
10^{-2}\ {\mathrm{GeV}}^4\ ,\\
\frac{\alpha_s}{\pi}\langle
GG\rangle\Big|_{\mathrm{Axial}}&=&(-2.2\pm0.4)\times 10^{-2}\ {\mathrm{GeV}}^4\ ,\nn
\end{eqnarray}
where the subscript denotes whether the condensate has been determined using the vector or the
axial-vector spectral function. Although these two quantities should be one and the same in QCD, it is
clear that the two values in Eq.~(\ref{gluonVA}) are not compatible with each other.  Of course,
the discrepancies (\ref{alphas}) and (\ref{gluonVA}) would not be there if
either a) one could argue that a specific analysis is to be preferred on
theoretical grounds, or b) if the errors
had been underestimated because of yet other systematic effects. Further determinations of the OPE condensates may be found in Refs. \cite{condensates:theworks}.

{}From the theoretical point of view, one could think of (at least) three sources of possible
systematic errors, namely the truncation of the perturbative series in powers of $\alpha_s$,
the truncation of the OPE, and the contribution from Duality
Violations (DVs). In fact, at some deeper level, the three sources must be related. First, the
purely perturbative contributions can be viewed as the contribution from the dimension-zero
``condensate'' ({\it i.e.}, the unit operator) to the OPE. Second, the lack of convergence of the
perturbative series in $\alpha_s$ calls for the existence of some nonperturbative contribution
which, through renormalons, is related to the OPE condensates. Finally, the lack of convergence of
the OPE is, in turn, at the origin of DVs.

It is far from clear that the different values for $\alpha_s$ quoted in Eqs.~(\ref{alphas}) should be due to DVs, since the first two error sources mentioned above can potentially explain this difference by themselves. However, in this work we will concentrate on DVs as a further possible source of error, because it has been much less explored  than the other two.

In QCD analyses of tau decays, the OPE plays a central role. However, although this
expansion is of such fundamental importance, the properties of the OPE in QCD are not known. In
particular,  although it is suspected that the OPE is asymptotic, this is not known to be true,
and, even if it is indeed asymptotic, we do not know whether it is summable or not, or how it
behaves along different rays in the $q^2$ complex plane. For an asymptotic expansion, it is of
course important to estimate its intrinsic theoretical error. Relegating a more precise
definition of this error to the discussion in the next section, here we just mention that this
intrinsic theoretical error is normally associated with DVs.

In practice, one can perhaps get a feeling of the systematic error in the perturbative expansion by
comparing different orders in the $\alpha_s$ expansion. Likewise, the systematic error in the
condensate contribution may also be assessed from the comparison between different orders in the
OPE. However, we do not know of any systematic approach for studying DVs. Therefore, although the
disagreement between, {\it e.g.},  the results quoted in Eq.~(\ref{gluonVA}) may point towards the fact that
``this method may approach its ultimate accuracy'' in tau decay~\cite{Davier}, there is currently
no method to estimate this accuracy from first principles in QCD. Hence, there appears to be no other way to make progress than to resort to models.

Based on tau data, the only attempt to date at estimating the error from DVs is the one made in
Ref.~\cite{Davier}; their claim is that DVs effects are completely negligible at the tau
mass. This result was obtained from the V+A spectral function with the help of a physically motivated model, previously used for
studying aspects of DVs~\cite{Shifman, Bigi, CGP1, CGP2}. In the present work, we reanalyze this
estimate studying separately the V and A correlators. As we will see, our conclusion is different: based on tau data alone, DVs may be larger than those found in Ref.~\cite{Davier}. The inclusion of $e^+e^-$ data beyond the $\tau$ mass, with
some extra assumptions to be detailed below, may also be used to further constrain the size of DVs but, even in this case, their impact could be comparable to the presently quoted systematic uncertainties. Therefore, the possible
effect of DVs should be taken into account.

This work is organized as follows. In Section~2, we present the theoretical
background needed for the discussion of DVs in tau decay, while in Section~3 we introduce and motivate our {\it ansatz}
 for DVs. The actual fits to the experimental tau data are described in Section~4.
Section~5 is devoted to an estimate of the impact of these fits on the
determination of $\alpha_s$ and, in Section~6, we assess the changes imposed on
this estimate if one also considers recent $e^+e^-$ data beyond the $\tau$ mass. Finally, in Section~7,
we offer some conclusions and prospects for future work.

\section{Duality violations and $\tau$ decay}
Let us start our discussion of hadronic tau decay by defining the correlators
\begin{eqnarray}\label{correlator}
\Pi^{V,A}_{\mu\nu}(q)&=&i\int\mathrm{d}^{4}x\,
e^{iq\cdot x}\langle \,0\,|\,
T\lbrace\, J_{\mu}(x)\,J_{\nu}^{\dagger}(0)\,\rbrace |\,0\,\rangle\,\nonumber\\
&=&(q^{\mu}q^{\nu}-q^2g^{\mu\nu})\,{\mathrm{Im}}\,\Pi_{V,A}^{(1)}(s)+
q^{\mu}q^{\nu}\,\Pi_{V,A}^{(0)}(q^2) \ ,
\end{eqnarray}
in which $J_{\mu}^V(x)=\bar{u}(x)\gamma_{\mu}d(x)$ and
$J_{\mu}^A(x)=\bar{u}(x)\gamma_{\mu}\gamma_5d(x)$. The ratio of the decay widths of
the tau lepton to nonstrange (vector and axial-vector channel) hadrons and the decay width to electrons, denoted as
\begin{equation}
R_{\tau}^{V,A}\equiv \frac{\Gamma[\tau^-\rightarrow (V^-,A^-)\nu_{\tau}]}{\Gamma[\tau^-\rightarrow
\nu_{\tau}e^-{\bar{\nu}}_e]}\ ,
\end{equation}
can be expressed as
\begin{equation}\label{spec}
R_{\tau}^{V,A}\!\!=12\pi
S_{EW}|V_{ud}|^2\!\!\! \int_0^{m_{\tau}^2}\!\! \frac{ds}{m_{\tau}^2}
\left(1-\frac{s}{m_{\tau}^2}\right)^2\left[\left(1+2\frac{s}{m_{\tau}^2}\right)
{\mathrm{Im}}\,\Pi_{V,A}^{(1)}(s)+{\mathrm{Im}}\,\Pi_{V,A}^{(0)}(s)\right]\ ,
\end{equation}
where the factor $S_{EW}=1.0201(3)$~\cite{EWrefs} accounts for small, known
electroweak corrections, and $V_{ud}$ is the corresponding entry in the
CKM matrix.

\begin{figure}
\renewcommand{\captionfont}{\small \it}
\renewcommand{\captionlabelfont}{\small \it}
\centering
\includegraphics[width=3in]{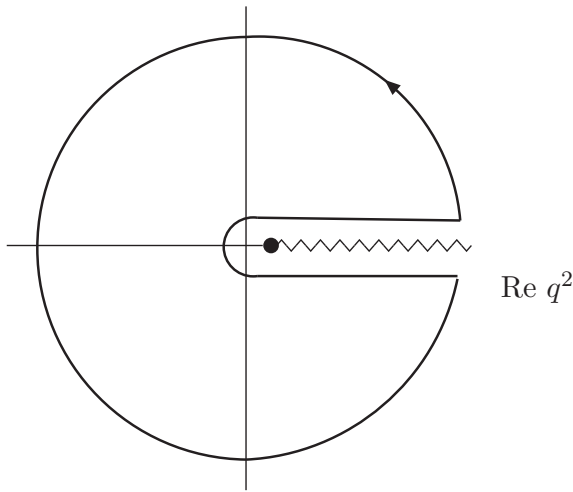}
\caption{Analytic structure of $\Pi_{V,A}(q^2)$ in the complex $q^2$ plane. The solid curve shows the contour used in Eq.~(\ref{cauchy1}).}\label{cauchy-fig}
\end{figure}

The spectral functions ${\mathrm{Im}}\,\Pi_{V,A}^{(1)}(s)$\footnote{${\mathrm{Im}}\,\Pi_{V}^{(0)}(s)$ vanishes in the isospin limit, which we assume,  and
${\mathrm{Im}}\,\Pi_{A}^{(0)}(s)$ is dominated by the pion pole.} appearing in Eq.~(\ref{spec})
have been
measured most accurately by ALEPH~\cite{Aleph}, and the results are included in
Fig.~\ref{aleph-fig} below. Since at present these spectral functions cannot
be directly calculated from QCD, what is being done in practice, following
Refs.~\cite{BNP, Shankar, FNdeR}, is to take advantage of the analytic properties of
the functions $\Pi_{V,A}^{(1)}(q^2)$ in the complex $q^2$ plane, depicted in
Fig.~\ref{cauchy-fig}. We write (dropping the superscript $^{(1)}$ from now on)
\begin{equation}\label{cauchy1}
\int_0^{s_0}\,ds\,\, P(s) \,\frac{1}{\pi}{\mathrm{Im}}\,\Pi_{V,A}(s)=-\frac{1}{2\pi i}\oint_{|q^2|=s_0}\,dq^2
\,P(q^2)\, \Pi_{V,A} (q^2)\ ,
\end{equation}
where $P(q^2)$ is an arbitrary polynomial which can be chosen at one's
convenience. In the application to tau decay one sets the radius $s_0=m_{\tau}^2$ and, under the assumption that this scale is larger than the scale at which the OPE sets in, it makes sense to
rewrite the right hand side of Eq.~(\ref{cauchy1}) as
\begin{equation}\label{cauchy2}
-\frac{1}{2\pi i}\oint_{|q^2|=s_0}\,dq^2
\,P(q^2)\, \Big\{\Pi^{ \mathrm{OPE}}_{V,A} (q^2)+   \Delta_{V,A}(q^2)\Big\}\ ,
\end{equation}
where, by definition,
\begin{equation}\label{dv1}
    \Delta_{V,A}(q^2)=\Pi_{V,A} (q^2)-\Pi^{ \mathrm{OPE}}_{V,A} (q^2)  \ ,
\end{equation}
and then approximate $ \Delta_{V,A}(q^2)\sim 0$, thus neglecting the second
term in Eq.~(\ref{cauchy2}). To the best of our knowledge, in all analyses of
tau decay to date, except for Ref. \cite{Davier}, this approximation has always been assumed. As we already mentioned in the introduction, Ref.~\cite{Davier} does discuss the possibility of DVs in tau decay, but only for the combined V+A correlator, and concludes that DVs are negligible and do not affect the systematic errors
from other sources, {\it cf.} the first result quoted in Eq.~(\ref{alphas}).\footnote{This conclusion will be reassessed in Section~5.}

Here we will not assume this approximation. On the contrary, we will take
nonvanishing contributions to $\Delta_{V,A}(q^2)$ as defined in Eq.~(\ref{dv1})
as the definition of DVs. Consequently, if we define
\begin{equation}\label{dv2}
{\cal{D}}_{V,A}^{[P]}(s_0)   =-\frac{1}{2\pi i}\oint_{|q^2|=s_0}\,dq^2
\,P(q^2)\, \Delta_{V,A}(q^2)\ ,
\end{equation}
as a measure of these DVs in tau decay, Eq.~(\ref{cauchy1}) becomes
\begin{equation}\label{cauchy3}
\int_0^{s_0}\,ds\,\, P(s) \,\frac{1}{\pi}{\mathrm{Im}}\,\Pi_{V,A}(s)=-\frac{1}{2\pi i}\oint_{|q^2|=s_0}\,dq^2 \,P(q^2)\, \Pi^{OPE}_{V,A} (q^2)+ {\cal{D}}_{V,A}^{[P]}(s_0) \  .
\end{equation}
Note that, if the OPE were a convergent expansion and $s_0$ were within the
radius of convergence, one would get $\Delta_{V,A}(q^2) =0$ and
thus DVs would be absent.

Although the precise convergence properties  of the OPE in the complex plane are not known, we do know that
this power expansion cannot be convergent but only asymptotic (at best). A convergent
expansion around $|q^2|=\infty$  in the complex plane defines
an analytic function on a complete annulus around the origin, which must,
therefore, also include the Minkowski axis. However, this is
contradicted by the existence of the physical cut along the Minkowski axis,
which shows  that the expansion cannot be convergent for any $s_0$.

In this work we will
assume that the OPE is an asymptotic expansion in $q^2$  and explore the
possible consequences. As an asymptotic expansion, the OPE will have a region of validity (the so-called wedge of asymptoticity \cite{Bender}) which will naturally exclude the Minkowski axis. This expectation is strongly supported by the large-$N_c$ limit,  for which the physical cut becomes an infinite set of poles which are not reproduced by the OPE.
We remark that, even if the function $P(q^2)$ is chosen such that it vanishes on the
Minkowski axis (as is the case for the polynomial appearing in the integrand of
Eq.~(\ref{spec})),
that does not guarantee that the DV function ${\cal{D}}_{V,A}^{[P]}(s_0) $ will exactly vanish.

\begin{figure}
\renewcommand{\captionfont}{\small \it}
\renewcommand{\captionlabelfont}{\small \it}
\centering
\includegraphics[width=3in]{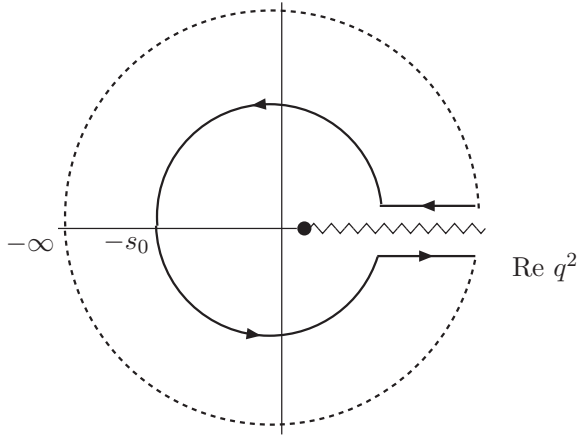}
\caption{Contour used in the derivation of
Eq.~(\ref{ourbaby}). With the assumed exponential decay, $\Delta_{V,A}(q^2)$ vanish on the
dashed circle with infinite radius.}\label{delta-fig}
\end{figure}

Accepting that the OPE is asymptotic, it is natural to assume a momentum
dependence of the DV function $\Delta_{V,A}(q^2)$ with an exponential decay at
large $|q^2|$.\footnote{This is the analog of the $e^{-1/\alpha}$-type error
one expects in an asymptotic expansion in powers of $\alpha$. Analyticity requires, however, the exponent to be angle dependent; see, {\it e.g.}, Sec.~III~B in Ref.~\cite{CGP1}.} Then,
using the contour depicted in Fig.~\ref{delta-fig} and the exponential decay of
$\Delta_{V,A}(q^2)$ on the dashed circle at infinity, one obtains a simpler expression
 for ${\cal{D}}_{V,A}^{[P]}(s_0) $ \cite{CGP1,CGP2}:
\begin{equation}\label{ourbaby}
    {\cal{D}}_{V,A}^{[P]}(s_0) =- \int_{s_0}^{\infty} ds\ P(s)\  \frac{1}{\pi}\mathrm{Im}\Delta_{V,A}(s)\ .
\end{equation}
 This relation is interesting for two reasons. It is useful, since
it expresses the DV contribution to Eq.~(\ref{cauchy3}),
$ {\cal{D}}_{V,A}^{[P]}(s_0)$, as an integral
 along  the Minkowski axis, in terms of the spectral function, and it can thus
in principle be determined from experimental data (as we will do below).
However, this requires an extrapolation from values of $s$ for which
experimental data are available
 to infinity. This shows explicitly the inherent
 difficulty present in any quantitative evaluation of DVs \cite{Donoghue}.

\section{An {\it ansatz} for Duality Violations}

 To make progress one clearly needs information about the functions $\mathrm{Im}\Delta_{V,A}(s)$. However, as emphasized above, it is not known how to get this information from QCD. Therefore, there is no other option but to adopt
 an  {\it ansatz} which is theoretically sensible and which is, of course,
not ruled out by existing data.  In other words, we look for a consistent picture of how duality violations
might occur in QCD, and  how they may affect the existing determinations of $\alpha_s$ (and,
consequently, also quark masses and  condensates). This way we are able to make an educated guess
about the possible size of the systematic errors due to DVs as allowed by existing data.

We will parametrize the duality-violating part of the spectral functions as
\begin{equation}\label{ansatz}
    \frac{1}{\pi}\mathrm{Im}\Delta_{V,A}(s)= \kappa_{V,A} \ e^{-\gamma_{V,A} s}\ \sin\left(\alpha_{V,A} +\beta_{V,A} s \right)\ ,
\end{equation}
The exponential decay in this expression is inspired
by our assumption that the OPE is an asymptotic expansion,
and can be understood as representing the finite width of resonances
\cite{Shifman}.
The oscillatory behavior is what
one expects in a spectral function with resonances exhibiting some kind of
periodicity.  The precise form chosen here is that obtained if
the vector and axial-vector resonances lie on Regge daughter trajectories.

\begin{figure}
\renewcommand{\captionfont}{\small \it}
\renewcommand{\captionlabelfont}{\small \it}
\centering
\includegraphics[width=2.9in]{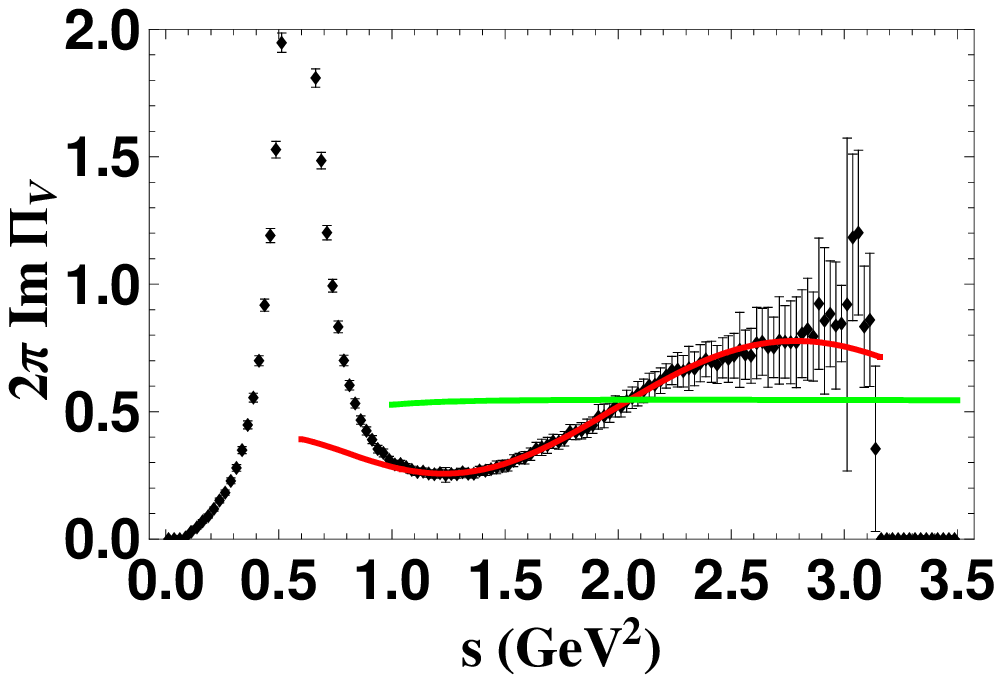}
\hspace{.1cm}
\includegraphics[width=2.9in]{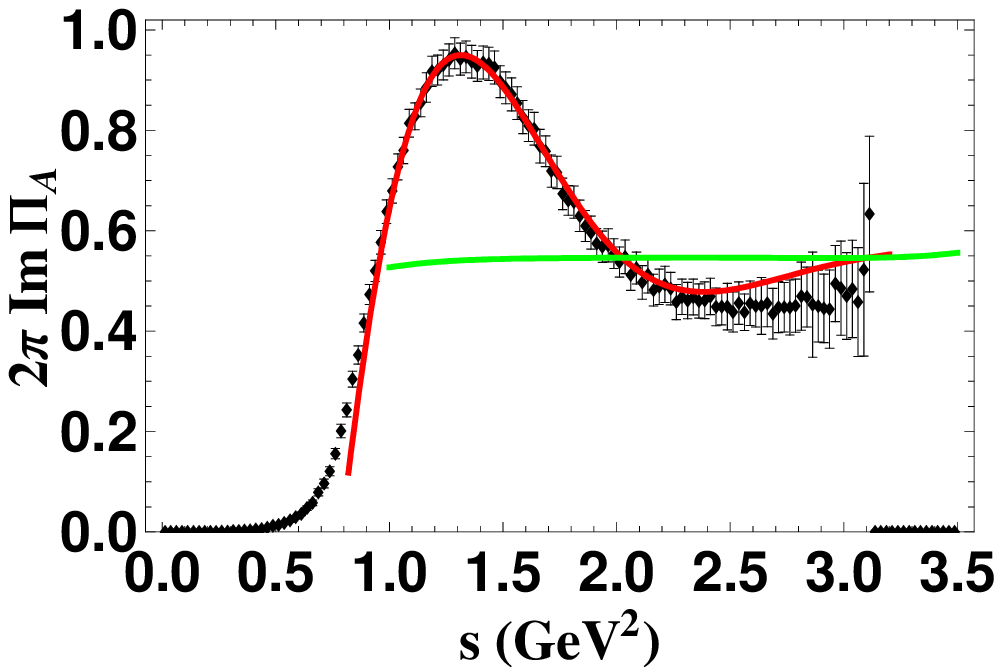}
\includegraphics[width=2.9in]{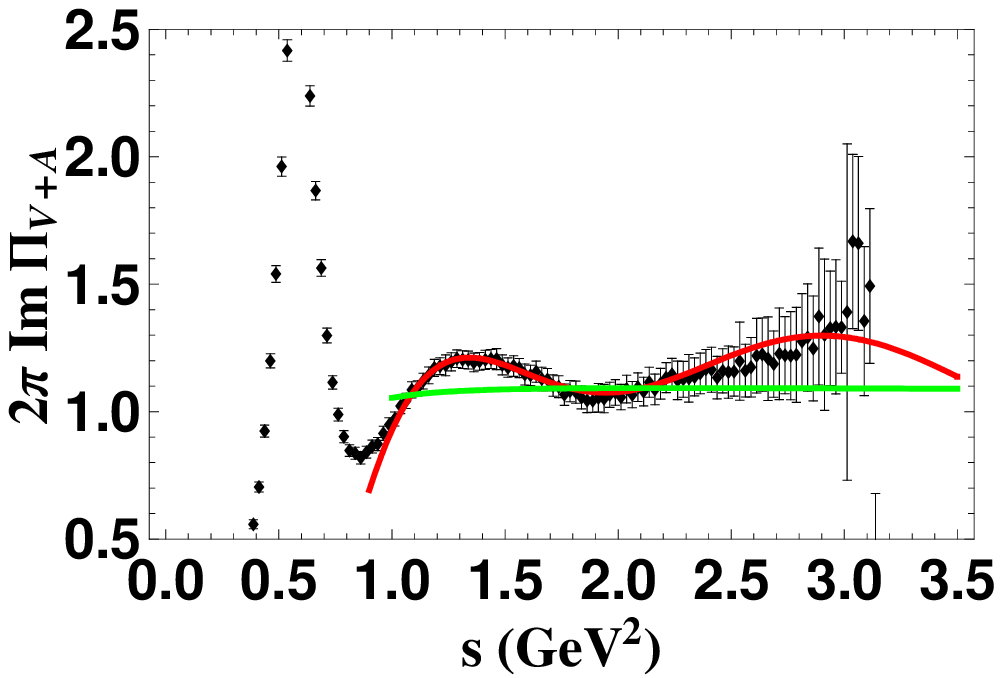}
\hspace{.1cm}
\includegraphics[width=2.9in]{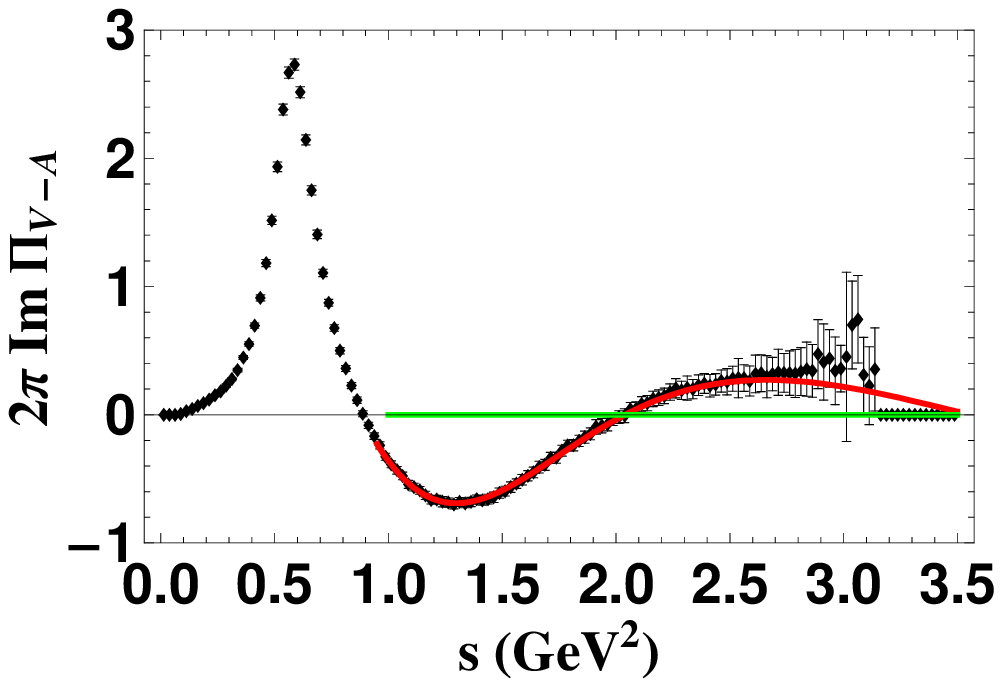}
\caption{Different combinations of spectral functions ($V$, $A$, and $V\pm A$)
(data points in black) compared with fixed-order perturbation theory
(green flat line) and the result of our fit to Eq.~(\ref{ansatz}), (red curve).}\label{aleph-fig}
\end{figure}

Furthermore, a glance at  Fig.~\ref{aleph-fig} reveals that both the
vector and the axial-vector spectral functions cross perturbation theory around $s=2$~GeV$^2$. A natural interpretation is that
the corresponding DV functions
$\mathrm{Im}\Delta_{V,A}(q^2)$ have a zero near this energy. Such a zero is much harder to understand without DVs
since, in this case, it would have to be due to a cancelation between different orders of the
OPE, signalling a possible breakdown of the OPE. Although we cannot
exclude such a breakdown at this rather high scale for minkowskian momentum, much evidence is consistent with the OPE being valid to a significantly lower scale for euclidean momentum, whether this scale is associated with the
gluon condensate or with the quark condensate. Such a hypothetical breakdown is also
not seen in the purely
perturbative contribution, which remains small and essentially flat all the way down to 1 GeV$^2$. Furthermore, as we will see in the next section and the Appendix, the OPE yields very small contributions to the $V$ and $A$ spectral functions  from the condensates of dimension four and six down to
$1$~GeV$^2$.

Since the width of a resonance is a $1/N_c$ effect, one expects the
exponent $\gamma$ to be relatively small.
The reason is that it has to go to zero when $N_c$ goes to infinity,  since
the exponential
suppression shown in Eq.~(\ref{ansatz}) has to disappear in favor of an
unsuppressed contribution representing  the isolated, infinitely narrow
peaks present in any spectral function in QCD in
the strict $N_c\rightarrow \infty$ limit. For the other parameters we expect $\kappa$ to be of order $N_c$, and
$\alpha$ and $\beta$ are expected to
be of order one in this limit.

Although our arguments are admittedly rather heuristic, it is reassuring that a mathematical model
realizing all the above features actually exists. In this model, first proposed in
Refs.~\cite{Shifman,Bigi}, resonance widths are introduced in a manner compatible with the analytic
properties of the correlators (\ref{correlator}), with masses following Regge theory. This model is
physically well motivated, and has been used for a variety of studies of DVs for both light and
heavy quarks~\cite{Bigi, CGP1,CGP2, Davier}.  Our {\it ansatz}, Eq.~(\ref{ansatz}), actually corresponds to the
asymptotic behavior of this mathematical model for large $s$. By considering the approximation (\ref{ansatz})
with {\it a priori} unknown values for the parameters $\alpha, \beta, \gamma$ and $ \kappa$, we
expect to describe the generic features of DVs at large $s$, avoiding some of the more model-dependent
details. We note, however, that the {\it ansatz} (\ref{ansatz}) is a very good approximation to the full
mathematical model already at relatively low values of $s$ \cite{CGP1}.

\section{Fit to tau data}

To find out whether the {\it ansatz} (\ref{ansatz}) is compatible with the tau experimental data, we have fitted the ALEPH spectral functions in both the vector and axial-vector channels, {\it i.e.}, we fitted the spectral data with the functions
\begin{equation}\label{impi}
    \frac{1}{\pi}\mathrm{Im}\Pi_{V,A}(s)= \theta(s-s_{min})\ \left\{\frac{N_c}{12 \pi^2}\big[ 1+ \widehat{\rho}(s) \big]+
    \kappa_{V,A} \ e^{-\gamma_{V,A} s}\ \sin\big(\alpha_{V,A} +\beta_{V,A} s \big)\right\} \ .
\end{equation}
The function $\widehat{\rho}(s)$ above contains the purely perturbative corrections in powers
of $\alpha_s$ up to  ${\cal{O}}(\alpha_s^5)$.\footnote{The term of ${\cal{O}}(\alpha_s^5)$ in not known exactly and we have used the  estimate provided by Ref. \cite{Jamin}.} These perturbative corrections differ depending on whether one uses CI or
FO perturbation theory and the expressions for both treatments  have been taken from
Ref.~\cite{Jamin}, where one can also find a discussion of their relevance for tau decay.

In principle one should also include in Eq.~(\ref{impi}) the contribution  from the condensates
which, away from $q^2=0$, can only come from the logarithms in the Wilson coefficients. However,
the contribution from the operators of dimension four and six turns out to be numerically very
small for reasonable values of the quark and gluon condensates. Our estimates for these
contributions are relegated to the Appendix. Condensates of dimension eight or higher will not be
considered.

As already mentioned in the previous section, the {\it{ansatz}} (\ref{ansatz}) is to be understood as an estimate of DVs
at large values of $s$. Since, in practice, we do not know how
large $s$ would have to be, we have done our fits in a window $s_{min}\leq s \leq
m_{\tau}^2$, varying the lower end, $s_{min}$. In other words, in the phenomenological approach we will be taking, the precise meaning of large values of $s$ will be set by the quality of the fit. The results of the fit are very
insensitive to the value of $s_{min}$, between a maximum value of $s_{min}$ of $\sim 1.8$ GeV$^2$ (above this value the
experimental errors are too large for the fit to be meaningful), and a minimum
value of $s_{min}$ of $\sim 1.1$ GeV$^2$. Below this value there is a sharp increase in the chi-squared per degree of freedom, showing a deterioration of
the fit, probably due to the tail of the $\rho$ meson (in the vector channel).
Consequently, we have chosen $1.1$ GeV$^2 \leq s \leq m_{\tau}^2$ as our
fitting window.

The result of the fit to the vector spectral function is then:
\begin{eqnarray}\label{fitV}
  \kappa_V &=& 0.018\pm 0.004 \ , \nn \\
  \gamma_V &=& 0.15\pm 0.15  \  \mathrm{GeV}^{-2}\ , \nn\\
  \alpha_V &=& 2.2\pm 0.3 \  ,\nn \\
  \beta_V &=& 2.0\pm 0.1 \   \mathrm{GeV}^{-2}\ , \nn \\
  \frac{\chi^2}{dof} &=& \frac{10}{79}\simeq 0.13\quad ,
\end{eqnarray}
and that to the axial-vector spectrum:
\begin{eqnarray}\label{fitA}
  \kappa_A &=& \ 0.20\pm 0.06  \ , \nn \\
  \gamma_A &=& \  1.7\pm 0.2  \ \mathrm{GeV}^{-2}\ , \nn\\
  \alpha_A &=& -0.4\pm 0.1 \ ,\nn \\
  \beta_A &=& -3.0\pm 0.1 \   \mathrm{GeV}^{-2}\ , \nn \\
  \frac{\chi^2}{dof} &=& \frac{17}{78}\simeq 0.22\quad .
\end{eqnarray}
Even though the perturbative term $\widehat{\rho}(s)$ in Eq.~(\ref{impi}) is
different in the CI or FO
resummation schemes, the results for the fit parameters are, within the quoted errors, insensitive to this
difference, or to the initial value for $\alpha_s$, which we took to be $\alpha_s(m_{\tau})=0.3156$. Our fits are also rather insensitive to an increase of the value
of $s_{min}$; the main effect is an increase of the errors quoted in
Eqs.~(\ref{fitV},\ref{fitA}).

Since our lower end is at $s=1.1$ GeV$^2$, one may  worry about the use of perturbation theory at such
low scales. However, there is nothing in the perturbative expansion signalling  a breakdown of the
approximation, as the nearly horizontal line in Fig. \ref{aleph-fig} clearly shows.  Also, one may
object that perturbation theory and the OPE should not be used for describing a spectral function on the
time-like axis, Eq. (\ref{impi}), {\it i.e.}, that ``local duality'' does not work.
However, we want to emphasize that local duality does not work \emph{precisely} because  DVs are not included, and that DVs, parametrized for example as in Eq.~(\ref{impi}), amount to adding
the necessary contributions which make up for the missing piece, therefore making local duality
exact ({\it cf.} Eq.~(\ref{cauchy3})).  Our analysis thus amounts to the assumption that
Eq.~(\ref{impi}) gives a reasonable account of DVs above $s=s_{min}$.

Looking at the values obtained from the fits (\ref{fitV},\ref{fitA}),
one immediately realizes that the oscillations in the axial-vector channel
will be more damped than in the vector channel as a result of a much larger
exponent, {\it i.e.}, $\gamma_A\gg \gamma_V$. While we have no understanding of why
this should be so in QCD, we simply observe that the difference in the values
of these parameters reflects the difference in the corresponding spectra as
measured by ALEPH.

 Fig. \ref{aleph-fig} shows the comparison of the fits to the different combinations
 of experimental spectral functions. As one can see, all fits are very good,
 having a very small
 $\chi^2/dof$. Our fits are fully correlated fits, {\it i.e.},
correlations among errors were taken into account in both
 fits using the pertinent covariance matrix provided by ALEPH.
 The small values of  $\chi^2/dof$ that we find
 might suggest an underestimation  of these correlations (or an overestimate of
 the diagonal errors), although it is difficult to know whether this is actually the
 case. In fact, an uncorrelated fit using only the diagonal entries of the covariance matrices
leads to results which are similar to those in Eq.~(\ref{fitV},\ref{fitA}), in particular for the vector spectral function.  We remark that our analysis is not the only one to get
 such small values for  $\chi^2/dof$. The analysis of Ref.~\cite{Davier} finds a value
 of  $\chi^2/dof=0.07$ in their OPE fit to moments such as Eq.~(\ref{spec}) in the vector channel.

As can be seen from Fig.~\ref{aleph-fig}, the vector and  axial-vector spectral functions oscillate
around perturbation theory in the window 1 GeV$^2\leq s\leq m_{\tau}^{2}$ with almost complete
anti-correlation, with a common``oscillation node'' around 2 GeV$^2$. Therefore the combination
$V+A$ agrees with perturbation theory better than $V$ or $A$ individually. Furthermore, the fits of
Refs.~\cite{BNP,Davier} suggest that there is a partial cancelation in the $V+A$ combination of the contribution
to the contour integral in Eq.~(\ref{cauchy2}) from both the dimension-six and dimension-eight
condensates in the OPE. These features have led people to believe that the combination $V+A$ is a
better correlation function
 than $V$ or $A$,  because it is somehow protected from all kinds of nonperturbative
 effects, including DVs. However, as the third panel on Fig.~\ref{aleph-fig} shows,
 the partial agreement of perturbation theory (green flat line) with the experimental
 $V+A$ data does not guarantee the absence of DVs. In fact, the results of our
 fits reported in Eqs.~(\ref{fitV},\ref{fitA}) reproduce the experimental data better than
 perturbation theory and, consequently, constitute a clear indication that nonzero
  DVs may exist in this spectral function as well. Furthermore, because DVs are suppressed
  in the axial-vector channel due to the  large exponent $\gamma_A$, the amount of DVs
  in $V+A$ is basically the same as in $V$. In our opinion, this example shows that one
   should understand the $V$ and $A$ channels individually before drawing any definite
    conclusion  about
 $V+A$.

\section{Impact of Duality Violations on $R_{\tau}$ and $\alpha_s$}

Once the DV functions (\ref{ansatz}), along with the values for the
parameters (\ref{fitV},\ref{fitA}), are known, one can use Eq.~(\ref{cauchy3})
to calculate the contribution from our DV {\it ansatz} to $R_{\tau}$. This is given by
 \begin{eqnarray}\label{impact}
   R_{\tau}^{V,A} &=&
(R_{\tau}^{V,A})^{(0)}
-  12 \pi S_{EW} |V_{ud}|^2 \int^{\infty}_{m_{\tau}^2} \frac{ds}{m_{\tau}^2} \left(1-\frac{s}{m_{\tau}^2}\right)^2 \left(1+2\frac{s}{m_{\tau}^2}\right) {\mathrm{Im}}\,\Delta_{V,A}(s) \nn \\
    &\equiv & (R_{\tau}^{V,A})^{(0)}
    + \delta  R_{\tau}^{V,A} \ ,
  \end{eqnarray}
 where $(R_{\tau}^{V,A})^{(0)}$ corresponds to the case of no DVs
 and, consequently,  $\delta  R_{\tau}^{V,A}$ corresponds to the contribution from
 duality violations. In the case of our {\it ansatz} (\ref{ansatz}), we obtain
 \begin{eqnarray}\label{deltaR0}
    \delta  R_{\tau}^{V,A}&=&-\ \frac{72\pi^2\kappa_{V,A}
    \ e^{-\gamma_{V,A} m_{\tau}^2}}{(\beta_{V,A}^2+\gamma_{V,A}^2)^4
    \ m_{\tau}^8} S_{EW}|V_{ud}|^2  \nn \\ &&\Big\{\Phi_1(m_{\tau}^2)
    \ \sin{\left(\alpha_{V,A}+\beta_{V,A}  m_{\tau}^2\right)}+
     \Phi_2(m_{\tau}^2)\ \cos{\left(\alpha_{V,A}+\beta_{V,A} m_{\tau}^2\right)}\Big\}\ ,
 \end{eqnarray}
 with
 \begin{eqnarray}\label{phis}
    \Phi_1(m_{\tau}^2) &=& 2 \beta_{V,A}^4-12\gamma_{V,A}^2\beta_{V,A}^2+2\gamma_{V,A}^4-
    \Big(3\gamma_{V,A}\beta_{V,A}^4+2\gamma_{V,A}^3\beta_{V,A}^2-
    \gamma_{V,A}^5\Big) m_{\tau}^2 \ ,\nn \\
   \Phi_2(m_{\tau}^2)  &=& 8\gamma_{V,A}\beta_{V,A}(\gamma_{V,A}^2-\beta_{V,A}^2)-
   \Big(\beta_{V,A}^5-2\gamma_{V,A}^2\beta_{V,A}^3-3\gamma_{V,A}^4\beta_{V,A}\Big)m_{\tau}^2\ ,
 \end{eqnarray}
for $s_0=m_\tau^2$.
Note that $ \delta  R_{\tau}^{V,A}$ goes to zero for $\gamma_{V,A}\rightarrow
\infty$, but is finite in the $\gamma_{V,A}\rightarrow 0$ limit. Upon substituting the values
from the vector and axial-vector fits (\ref{fitV},\ref{fitA}) one finds
\begin{equation}\label{deltaR}
\delta R_{\tau}^V=\left( -0.022\pm 0.013\right)\ S_{EW}|V_{ud}|^2\  ;\  \delta R_{\tau}^A=\left(0.0003\pm 0.0003\right)\ S_{EW}|V_{ud}|^2\ ,
\end{equation}
which translates into $\delta R_{\tau}^V/R_{\tau}^V\simeq -(1.5\pm 0.9)\%$ and $\delta R_{\tau}^A/R_{\tau}^A\simeq (0.02\pm 0.02)\%$. The effect of DVs is much smaller in the axial-vector channel,
because of the larger $\gamma_A$, as anticipated.

For comparison, Ref.~\cite{Davier} takes the resonance model in Ref.~\cite{Shifman} as a model for DVs, matches it directly to the $V+A$ spectral function near  $s=m^2_{\tau}$, and finds that the effect of DVs is at most $\delta R_{\tau}^{V+A}\simeq 0.0021 S_{EW}|V_{ud}|^2$. In our case we keep only the form of the asymptotic behavior of this type of model for the spectral function ({\it cf}. Eq.~(\ref{ansatz})), and we then fit the parameters of this {\it ansatz} to the
$V$ and $A$ spectral functions separately. We thus find for the effect in $V+A$ the sum of the two results in
Eq.~(\ref{deltaR}),  which is ten times larger than the estimate of Ref.~\cite{Davier}, but still consistent with the experimental data. Ref.~\cite{Davier} also considers an instanton-based model \cite{Shifman}. In this case, the model does not have the exponential suppression of Eq.~(\ref{ansatz}), and the contribution becomes  $\delta R_{\tau}^{V+A}\simeq 0.014 S_{EW}|V_{ud}|^2$. As it turns out,  this is closer to our estimate in magnitude, although opposite in sign.

In order to estimate the shift in $\alpha_s$ induced by these values for $\delta R_{\tau}^{V,A}$, we first note that the perturbative expression for $R_{\tau}^{V+A}$ can be written as
\begin{equation}\label{Rtaupert}
    (R_{\tau}^{V+A})_{PT}=\ N_c\ S_{EW} |V_{ud}|^2 \left [ 1+ \delta_{PT}\right]\ ,
\end{equation}
with  $\delta_{PT}$ given by the following expansions \cite{Jamin}
\begin{eqnarray}\label{PT}
  \delta_{PT}^{FO}& = &  0.1082 \left(\frac{\alpha_s(m_{\tau})}{0.34}\right) +  0.0609  \left(\frac{\alpha_s(m_{\tau})}{0.34}\right)^2 +\nn \\
 &&  0.0334  \left(\frac{\alpha_s(m_{\tau})}{0.34}\right)^3 + 0.0174  \left(\frac{\alpha_s(m_{\tau})}{0.34}\right)^4 +  0.0088  \left(\frac{\alpha_s(m_{\tau})}{0.34}\right)^5 \ , \\
  \delta_{PT}^{CI} &=&  0.1479 \left(\frac{\alpha_s(m_{\tau})}{0.34}\right) +  0.0297  \left(\frac{\alpha_s(m_{\tau})}{0.34}\right)^2 +  \nn \\
 && 0.0122  \left(\frac{\alpha_s(m_{\tau})}{0.34}\right)^3
  + 0.0086  \left(\frac{\alpha_s(m_{\tau})}{0.34}\right)^4 +  0.0038  \left(\frac{\alpha_s(m_{\tau})}{0.34}\right)^5   , \\ \nn
\end{eqnarray}
depending on whether one uses the CI or FO  prescription.\footnote{The last term in these expansions is an estimate of the $\mathcal{O}(\alpha_s^5)$ term from Ref.~\cite{Jamin}.} Using the experimental value for the tau decay ratio $R_{\tau}^{V+A}$ plus a conservative estimate for the main contributions from the condensates, Ref.~\cite{Jamin} estimates  the phenomenological value for the parameter $\delta$ as
\begin{equation}\label{delta}
    \delta_{phen}= 0.2042\pm 0.0050\ .
\end{equation}
Equating $1+ \delta_{PT}+ \delta R_{\tau,V+A}/R_{\tau,V+A}$ to $1+\delta_{phen}$, one obtains an estimate for the  shift in $\alpha_s$ due to the DV contribution (\ref{deltaR0}), yielding
\begin{equation}\label{shift}
  \delta \alpha_s(m_{\tau}) \sim  0.003-0.010 \ ,
 \end{equation}
both for CI and FO prescriptions. The spread of values in Eq.~(\ref{shift}) reflects the error in the sum of the $V$ and $A$ results in  Eq.~(\ref{deltaR}), {\it i.e.}, in $ \delta R_{\tau}^{V+A}$.

We consider our result, Eq.~(\ref{shift}), as a fair estimate of the systematic error  associated with duality violations in $\alpha_s(m_{\tau})$, as determined from the total nonstrange tau decay width. However, the value of $\alpha_s(m_{\tau})$ and the condensates may also be determined from a combined fit to a set of moments written in terms of pinched weights (see, {\it e.g.}, \cite{Davier, Maltman}). In this case, our estimate is probably not good enough and a full analysis of these moments in the presence of DVs is required, for instance along the lines of our recently proposed iterative method described in Ref.~\cite{CGP1}. At any rate, it is clear that the error associated with DVs is not numerically negligible, at least in the vector channel.

\section{Inclusion of $e^{+}e^{-}$ data}

Since, according to Eq.~(\ref{ourbaby}),  DVs entail
an extrapolation to high energies and, with our  {\it ansatz},
they turn out to be sizeable in the vector channel,
it makes sense to ask whether $e^+e^-$ data, which extend
to higher energies than tau data, can be used to further restrict the range of values for the DV
 parameters (\ref{fitV}).

However, one cannot relate DVs in tau decay to those in $e^+e^-$ without further assumptions. Since
tau data concerns currents with a flavor structure of the type $\overline{u}d$, whereas $e^+e^-$
data see the flavor-singlet combination\footnote{We will restrict our discussion to energies below
the $\overline{c}c$ spectrum.} $(2/3) \overline{u}u -(1/3) \overline{d}d -(1/3) \overline{s}s$,
there is an OZI suppressed contribution in $e^+e^-$ which is absent in tau decay. Although this
contribution is $1/N_c$ suppressed and, in perturbation theory, shows up only
at $\mathcal{O}(\alpha_s^3)$, it is not clear how much it might affect the value of the DV parameters
appearing in Eq.~(\ref{ansatz}), since they encode nonperturbative effects on the
time-like axis.
As an exploratory step, we will simply assume that OZI
contributions do not significantly affect the values of the parameters in the DV function (\ref{ansatz}).
This is not the only difficulty, however. The strange quark mass is much larger than the up
and down masses, and this should affect the values of the DV parameters for the
$\overline{s}s$ component. Fortunately, taking guidance from the model
 underlying the form of our {\it ansatz} \cite{Shifman, Bigi, CGP1,CGP2}, one finds that,
 while the parameters $\kappa_V, \gamma_V$ and $ \beta_V$ are related to the universal slope of the
 daughter Regge trajectories, the value of $\alpha_V$ is related to the mass (squared) of the
 lowest~lying resonance of the Regge tower. This suggests that, while the value of $\alpha_V$ should
 be similar in the $\overline{u}u$ , $\overline{d}d$ and $\overline{u}d$ channels (assuming
isospin symmetry),
 in the $\overline{s}s$ channel it should be shifted by a certain amount proportional to the strange
  quark mass. Since we do not know how large this shift may be, in practice we introduce a
  new parameter $\alpha'_V$ for the  $\overline{s}s$ component.

\begin{figure}
\renewcommand{\captionfont}{\small \it}
\renewcommand{\captionlabelfont}{\small \it}
\centering
\includegraphics[width=2.9in]{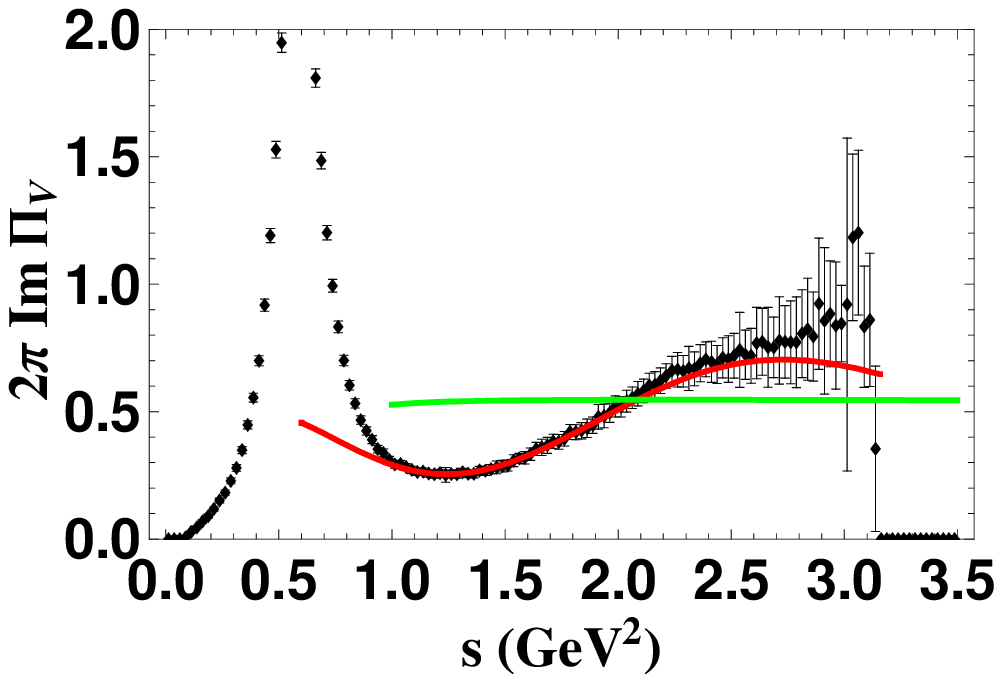}
\hspace{0.1cm}
\includegraphics[width=2.9in]{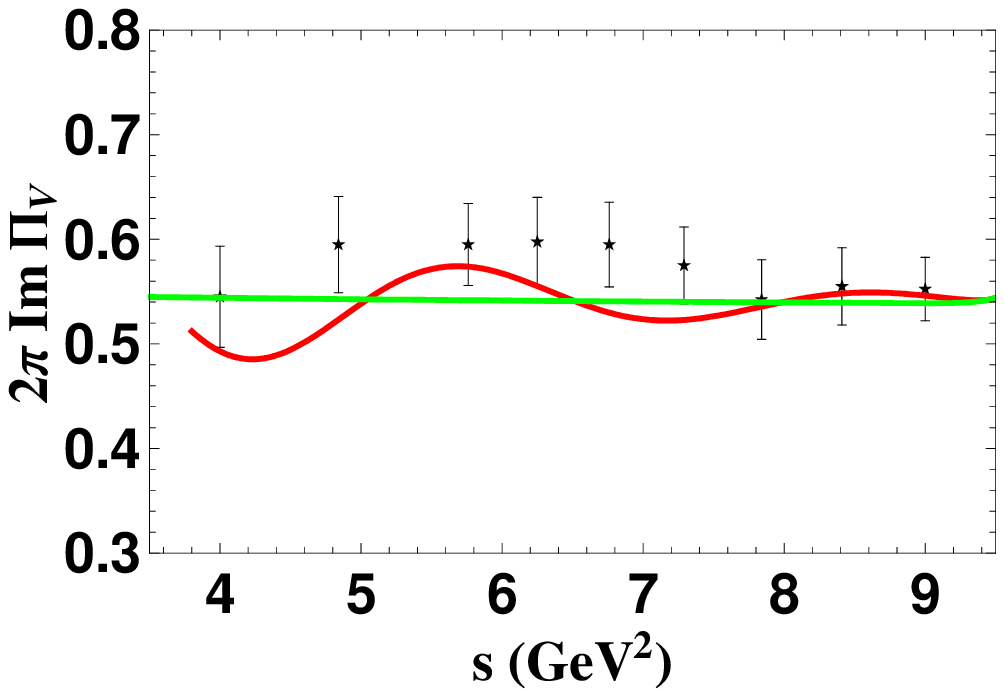}
\caption{Plots of the vector spectral function in tau decay (left panel) and $e^+e^-$ (right panel) as compared to perturbation theory (green flat line) and the result of the combined fit (\ref{hybridfit}) (red oscillating curve). }\label{epem-fig}
\end{figure}

Therefore, in the context of $e^+e^-$ data, the spectral function including the {\it ansatz} for DVs
becomes\footnote{Because of OZI-suppressed contributions, the function
$\widehat{\rho}(s)$ in $e^+e^-$ differs from that in tau decay at $\mathcal{O}(\alpha_s^3)$, as we
have already mentioned. However, the result of the fit is insensitive to these perturbatve details,
and we will not take them into account.}
\begin{eqnarray}\label{impiee}
      \frac{1}{\pi}\mathrm{Im}\Pi_{V}(s)&=&\frac{N_c}{12 \pi^2}\left[ 1+ \widehat{\rho}(s) \right]+\nn\\
         && \frac{5}{6}\ \kappa_{V} \ e^{-\gamma_{V} s}\ \sin\left(\alpha_{V} +\beta_{V} s \right) + \frac{1}{6}\ \kappa_{V} \ e^{-\gamma_{V} s}\ \sin\left(\alpha'_{V} +\beta_{V} s \right)\ ,
\end{eqnarray}
where the new parameter $\alpha'_{V}$ takes  this aforementioned
shift into account, and the weights 5/6 and 1/6 correspond to the
sum of the squares of nonstrange and strange electric charges, after the spectral function
is normalized as in tau decays, {\it cf.} Eq.~(\ref{impi}).

We have attempted a fit of Eq. (\ref{impiee})  to the $e^+e^-$
data\footnote{These data are available at http://pdg.lbl.gov/current/xsect/.}
in the window $1.1$ GeV$^2\lesssim s_{\min}\lesssim 9$ GeV$^2$ but, as it turns out,
the quality of these data is not good enough to allow a stable result for this fit.
Another issue of concern is that the $e^+e^-$ points come from a
compilation of different experiments, each with its own systematics, and it is not clear to us how to take the systematic errors properly into account (see Ref.~\cite{Eidelman:1995ny} for a discussion).

Therefore, we decided to proceed as follows. Since the problem with tau data is
that they do not extend above $s=m^2_{\tau}$, we have supplemented these data
with the spectral function from $e^+e^-$ in the range  4 GeV$^2\lesssim s_{\min}\lesssim 9$ GeV$^2 $
as provided by BES \cite{Bai:2001ct}, at present the most precise (inclusive) measurement in this energy range.
Inconsistencies between exclusive and inclusive determinations of the $e^+e^-$ cross section in the
region 3 GeV$^2\leq s\leq 4$ GeV$^2$ have prevented us from using these data as well.  This is a
longstanding issue (see, for instance, Ref.~\cite{Eidelman:1995ny}) which, to the best of our knowledge,
 has not yet been fully resolved. Because of this, and because of all the extra assumptions we had to make
 to arrive at Eq.~(\ref{impiee}), we emphasize that the present exercise should be
 seen only as a very preliminary estimate of constraints from $e^+e^-$ data,
 rather than as a full-fledged analysis. For a more reliable analysis
 a better understanding of the theoretical aspects  and better data in the $2-4$ GeV$^2$ window would be needed. The
 latter may soon be provided by BaBar \cite{Babar}.

With all these caveats in mind, we are now ready to address the question as to whether the
values of the DV parameters in Eq.~(\ref{fitV}) are consistent
with $e^+e^-$ data above $s=m^2_{\tau}$.  To try answer this question, we have performed a
combined fit in which we have taken the {\it ansatz} (\ref{impi}) to fit
 the tau data in the region $1.1$ GeV$^2\lesssim s_{\min}\lesssim m^2_{\tau}$ and,
 simultaneously,  the {\it ansatz} (\ref{impiee}) to fit the $e^+e^-$ data in the
 region $4$ GeV$^2\lesssim s_{\min}\lesssim 9$ GeV$^2$. The result of this combined fit
  yields the following values:
\begin{eqnarray}\label{hybridfit}
  \kappa_V &=& 0.024\pm 0.004 \nn \\
  \gamma_V &=& 0.40 \pm 0.12 \nn \\
  \alpha_V &=& 1.82 \pm 0.19\nn \\
  \beta_V &=& 2.14\pm 0.11\nn \\
  \alpha'_V &=& 5.2\pm 1.4 \nn \\
  \frac{\chi^2}{dof} &=& \frac{22}{87}\simeq 0.25  \ .
\end{eqnarray}
 As one can see, our {\it ansatz} produces a reasonable fit to the data with values for
 the parameters which are compatible, within errors, with the results in Eq.~(\ref{fitV}).
  There is a certain trend towards smaller DVs, in comparison with our results from
tau data alone.  We note that if we compute the $\chi^2$ with the parameter values
given in Eq.~(\ref{hybridfit}) using only the (nine) BES data points, we find a value
$\chi^2/dof = 10/9$.
   The comparison of the fits with the data, and with perturbation theory,
   can be seen in Fig.~\ref{epem-fig}.

Using the values (\ref{hybridfit}), we can estimate the DVs by substitution into
Eq.~(\ref{deltaR0}), which yields
\begin{equation}\label{newRtau}
    \delta R_{\tau}^V= -(0.0065\pm 0.0042) \ S_{EW}|V_{ud}|^2\  .
\end{equation}
Following the same steps leading to Eq.~(\ref{shift}), the corresponding shift in $\alpha_s$, as determined from the tau decay width, is given by
\begin{equation}\label{newalpha}
    \delta \alpha_s(m_{\tau}) \sim  0.001-0.003 \ .
\end{equation}
This value is roughly a factor three smaller than that previously found in the fit to tau data alone,
 Eq.~(\ref{shift}), but it is not negligibly small in comparison with  the theoretical
errors quoted in Eq.~(\ref{alphas}).

\section{Conclusions and outlook}

In this work we have considered the problem of evaluating the possible size of DVs in tau decay.
 Even though there are good reasons to believe that DVs are present in QCD, our lack of
 understanding of these effects makes it very difficult to incorporate them in a systematic
 analysis of the data. However, ignoring the issue altogether may lead to a underestimation
 of systematic effects in the extraction of
 $\alpha_s$, quark masses and condensates. Some estimate of the
effect of DVs is needed, given  the high level of accuracy presently sought.

A general framework for the contribution of DVs is provided by Eqs.~({\ref{cauchy3},\ref{ourbaby}).
In order to get to a quantitative estimate of DVs, however, more detailed information about the functions
$\mathrm{Im}\Delta_{V,A}(s)$ is needed.   In order to do this, we proposed an
{\it ansatz} (\ref{ansatz}) which is based on
the models of Refs.~\cite{Shifman,Bigi,CGP1,CGP2}, but which we expect to be of
more general validity. Although this validity is speculative, at least it has the
virtue of being falsifiable with future improvements in experimental data.

At present, the {\it ansatz} is perfectly
 compatible with the tau data, producing good fits for both the vector and the
 axial-vector spectral functions, and allowing us to determine
 the parameters of the {\it ansatz}, {\it cf.} Eqs.~(\ref{fitV},\ref{fitA}). These
parameter values lead to
 the estimate (\ref{deltaR}) for the contribution from DVs to $R_{\tau}$,
 and, from that, to a possible theoretical error in $\alpha_s(m_{\tau})$
 of order 0.003--0.010. Since we find good fits for both vector and axial-vector
channels, our fits also give a good description of the $V+A$ spectral function.
However, we find contributions to $R_{\tau}$ which are much larger than those based on a similar model in Ref.~\cite{Davier}, which were obtained from an estimate based on the $V+A$ spectral function only.

Inclusion of $e^+e^-$ data, plus some assumptions, modifies the
 values of the parameters in the {\it ansatz} to those given in (\ref{hybridfit}).
  These modified values lead to a reduction of the effect of DVs on $R_\tau$. This is just a
  consequence of the flatness of the inclusive $e^+e^-$ data above 4 GeV$^2$ \cite{Bai:2001ct},
to which we have limited ourselves here,
  as they appear to be more reliable than the exclusive data at lower
  energies. However, given that  exclusive data below 4 GeV$^2$ are not
  consistent with this behavior, it will be important to clarify this issue,
  once more and better data become available. At any rate, even with the inclusion of the
$e^+e^-$ data of Ref.~\cite{Bai:2001ct},
  we find that DVs may lead to a systematic error which is as large as
  $\delta\alpha_s(m_{\tau})|_{th}\simeq 0.003$, {\it i.e.}, of the order of half
  the systematic errors from all other sources taken
   together in \cite{Davier, Jamin}, {\it cf.}  Eq.~(\ref{alphas}). In a conservative approach,
    systematic errors with different origin ought to be added
    linearly. In light of these results, we conclude that
    future QCD analyses based on tau data should no longer neglect the possible contributions from  DVs.

\vspace{1cm}

\textbf{Acknowledgements}\\

We thank Matthias Jamin for his collaboration at different stages of the present work, as well as for discussions.
We would also like to thank Michel Davier, Sebastien Descotes-Genon, Andreas H\"ocker, Louis Lyons, Bogdan Malaescu, Kim Maltman, Ram\'on Miquel, Lluisa Mir, Antonio Pich, Eduardo de Rafael and Graziano Venanzoni for useful discussions.  MG would like to thank IFAE at the UAB for hospitality. This work was supported in part by CICYT-FEDER-FPA2005-02211, SGR2005-00916, the Spanish Consolider-Ingenio 2010 Program CPAN (CSD2007-00042), by the EU Contract No. MRTN-CT-2006-035482, ``FLAVIAnet,'' and by the US Department of Energy.

\vspace{2cm}
 \renewcommand{\theequation}{A.\arabic{equation}}
  \setcounter{equation}{0}  
  \section*{Appendix: Estimate of OPE contributions to ${\mathrm{Im}}\,\Pi_{V,A}$}  

\vspace{1cm} In the analysis of Sec.~4, the analytic form of the {\it ansatz} used to fit the
spectral
function, Eq.~({\ref{impi}), assumed that the contribution from the OPE to the imaginary part of the correlator is  negligible. In this appendix we will estimate the imaginary
contributions  coming from ${\cal{O}}(\alpha_s)$ corrections from dimension four and six operators.

We start by splitting the vector and axial correlators as
\begin{equation}
\Pi_{V,A}=\Pi_{V,A}^{(0)}+\Pi_{V,A}^{(2)}+\Pi_{V,A}^{(4)}+\Pi_{V,A}^{(6)}\ ,
\end{equation}
where the superscripts refer to the dimension of the OPE terms. The perturbative contributions are given by~\cite{BNP}
\begin{equation}
\Pi_V^{(0)}=\Pi_A^{(0)}=-\frac{1}{4\pi^2}\log{\frac{Q^2}{\mu^2}}\left[1+\frac{\alpha_s(Q)}{\pi}+\cdots\right]\ .
\end{equation}
Contributions from $\Pi_{V,A}^{(2)}$ are proportional to the $u,d$ quark masses and can be safely neglected. The contribution from dimension four operators is, from Ref.~\cite{BNP}
\begin{eqnarray}
Q^4\, \Pi_V^{(4)}& =& \frac{1}{12}\left[1-\frac{11}{18}\frac{\alpha_s(Q)}{\pi}+
\cdots\right]\langle\frac{\alpha_s}{\pi}G_{\mu\nu}G^{\mu\nu}\rangle \nn\\
&&+\left[1+\frac{13}{27}\frac{\alpha_s(Q)}{\pi}+
\cdots\right](m_u+m_d)\langle{\bar{\psi}}\psi\rangle+ \frac{4}{27} \frac{\alpha_s(Q)}{\pi} \ m_s\langle{\bar{\psi}}\psi\rangle\;,\nonumber\\
Q^4\, \Pi_A^{(4)}& =& \frac{1}{12}\left[1-\frac{11}{18}\frac{\alpha_s(Q)}{\pi}+
\cdots\right]\langle\frac{\alpha_s}{\pi}G_{\mu\nu}G^{\mu\nu}\rangle  \\
&&+\left[1-\frac{59}{27}\frac{\alpha_s(Q)}{\pi}+
\cdots\right](m_u+m_d)\langle{\bar{\psi}}\psi\rangle + \frac{4}{27} \frac{\alpha_s(Q)}{\pi} \ m_s\langle{\bar{\psi}}\psi\rangle\  ,\nn
\end{eqnarray}
whereas for dimension six operators one finds \cite{BNP}
\begin{eqnarray}
Q^6 \, \Pi_V^{(6)}&=&-8\pi^2\left[1-\frac{91}{72}\log{\frac{Q^2}{\mu^2}}\left(\frac{\alpha_s(\mu)}{\pi}\right)+\cdots\right]\left(\frac{\alpha_s(\mu)}{\pi}\right)\langle A_{\mu}^a A^{\mu\,a}\rangle\nonumber\\
&&-\frac{32\pi^2}{9}\left[1-\frac{643}{288}\log{\frac{Q^2}{\mu^2}}\left(\frac{\alpha_s(\mu)}{\pi}\right)+\cdots\right]\left(\frac{\alpha_s(\mu)}{\pi}\right)\langle V_{\mu}^a V^{\mu\,a}\rangle\nonumber\\
&&+\frac{16\pi^2}{27}\left[\log{\frac{Q^2}{\mu^2}}\left(\frac{\alpha_s(\mu)}{\pi}\right)^2+\cdots\right]\langle A_{\mu} A^{\mu}\rangle\nonumber\\
&&+
\frac{8\pi^2}{3}\left[\log{\frac{Q^2}{\mu^2}}\left(\frac{\alpha_s(\mu)}{\pi}\right)^2+\cdots\right]\langle V_{\mu} V^{\mu}\rangle\ , \nn
\end{eqnarray}
\begin{eqnarray}
Q^6 \, \Pi_A^{(6)}&=&\frac{55\pi^2}{9}\left[\log{\frac{Q^2}{\mu^2}}\left(\frac{\alpha_s(\mu)}{\pi}\right)^2+\cdots\right]\langle A_{\mu}^a A^{\mu\,a}\rangle\nonumber\\
&&-\frac{88\pi^2}{9}\left[1-\frac{319}{792}\log{\frac{Q^2}{\mu^2}}\left(\frac{\alpha_s(\mu)}{\pi}\right)+\cdots\right]\left(\frac{\alpha_s(\mu)}{\pi}\right)\langle V_{\mu}^a V^{\mu\,a}\rangle\nonumber\\
&&+\frac{88\pi^2}{27}\left[\log{\frac{Q^2}{\mu^2}}\left(\frac{\alpha_s(\mu)}{\pi}\right)^2+\cdots\right]\langle A_{\mu} A^{\mu}\rangle\ ,
\end{eqnarray}
where in all the equations above $(\cdots)$ stands for higher order contributions in the strong coupling constant.
In the dimension six contribution we have kept only the leading log contribution (finite pieces will not contribute to the imaginary part), and we used the following short-hand notation for the four-quark condensates:
\begin{eqnarray}
\langle A_{\mu}^a A^{\mu\,a}\rangle&=&\langle {\bar{\psi}}\gamma_{\mu}\gamma_5 T^a \psi{\bar{\psi}}\gamma^{\mu}\gamma_5 T^a\psi\rangle\ ,\nonumber\\
\langle V_{\mu}^a V^{\mu\,a}\rangle&=&\langle {\bar{\psi}}\gamma_{\mu} T^a \psi{\bar{\psi}}\gamma^{\mu} T^a\psi\rangle\ ,\nonumber\\
\langle A_{\mu} A^{\mu}\rangle&=&\langle {\bar{\psi}}\gamma_{\mu}\gamma_5 \psi{\bar{\psi}}\gamma^{\mu}\gamma_5\psi\rangle\ ,\nonumber\\
\langle V_{\mu} V^{\mu}\rangle&=&\langle {\bar{\psi}}\gamma_{\mu} \psi{\bar{\psi}}\gamma^{\mu}\psi\rangle\ .
\end{eqnarray}
Assuming vacuum factorization \cite{SVZ}, the four-quark condensates simplify to
\begin{eqnarray}
&&\langle A_{\mu}^a A^{\mu\,a}\rangle=-\langle V_{\mu}^a V^{\mu\,a}\rangle=\frac{4}{9}\langle{\bar{\psi}}\psi\rangle^2\ ,\nonumber\\
&&\langle A_{\mu} A^{\mu}\rangle=-\langle V_{\mu} V^{\mu}\rangle=\frac{2}{9}\langle{\bar{\psi}}\psi\rangle^2\ .
\end{eqnarray}
In order to calculate ${\mathrm{Im}}\,\Pi_{V,A}(t)$, one uses that
\begin{equation}
\alpha_s(Q)=\alpha_s(\mu)\left\{1+\frac{\beta_1}{2}\frac{\alpha_s(\mu)}{\pi}\log{\frac{Q^2}{\mu^2}}\right\}\ ,
\end{equation}
with
\begin{equation}
\beta_1=\frac{2n_f-11N_c}{6}\Bigg|_{n_f=3}=-\frac{9}{2}\ .
\end{equation}
As a result,
\begin{equation}
{\mathrm{Im}}\,[\alpha_s(Q)]=\frac{9}{4}[\alpha_s(Q)]^2\ .
\end{equation}
It is now straightforward to calculate the different OPE contributions to the spectral functions.
For normalization purposes it is convenient to work with $v_1\equiv 2\pi {\mathrm{Im}}\,\Pi_V$, $a_1\equiv 2\pi {\mathrm{Im}}\,\Pi_A$ . Taking $\alpha_s(1.1\,{\mathrm{GeV^2}}) / \pi\simeq 0.15$, one finds for the perturbative contribution:
\begin{equation}\label{pert}
\delta v_1^{(0)}=\delta a_1^{(0)}=\frac{1}{2}\left[1+\frac{\alpha_s(Q)}{\pi}\right]\simeq 0.58\ .
\end{equation}
Taking $\langle \alpha_s G^2 \rangle \simeq 0.01 \pi \,\,{\mathrm{GeV}}^4$, the gluon condensate contribution amounts to
\begin{equation}
\delta v_1^{(G^2)}=\delta a_1^{(G^2)}\simeq -4\cdot 10^{-4}\,\, \frac{{\mathrm{GeV}}^4}{Q^4}\ .
\end{equation}
For the quark condensate contribution, we will assume SU(3) symmetry in the condensates, the PCAC relation  $(m_u+m_d)\langle {\bar{\psi}}\psi \rangle=-m_{\pi}^2f_{\pi}^2$,  and estimate $m_s/(m_{u}+ m_{d})\sim 10$. This yields
\begin{equation}
\delta v_1^{(\langle {\bar{\psi}}\psi\rangle)}\simeq -3\cdot 10^{-4}\,\, \frac{{\mathrm{GeV}}^4}{Q^4}\ ,\ \ \ \ \
\delta a_1^{(\langle {\bar{\psi}}\psi\rangle)}\simeq 1\cdot 10^{-4}\,\, \frac{{\mathrm{GeV}}^4}{Q^4}\ .
\end{equation}

In Ref.~\cite{BNP} it was argued that the biggest non-perturbative contributions to $R_{\tau}$ come from the dimension six operators. There is some  evidence that naive factorization does not work for these condensates. Based on a phenomenological fit, Ref.~\cite{BNP} concluded that there might be an enhancement over the result from factorization that can be represented by rescaling the condensate
\begin{equation}
\langle{\bar{\psi}}\psi\rangle^2 \longrightarrow \lambda \langle{\bar{\psi}}\psi\rangle^2 \ ,
\end{equation}
where $\lambda$ is a fudge factor which, for typical values of $\alpha_s$ and the quark condensate, can be estimated to be $\lambda\simeq 3-9$.  We take the quark condensate $\langle {\bar{\psi}}\psi \rangle = (-240\,{\mathrm{MeV}})^3$ and, to be conservative, we take $\lambda=10$. Accordingly,
\begin{equation}
\delta v_1^{(6,\lambda)}\lesssim 4\cdot 10^{-3}\,\, \frac{{\mathrm{GeV}}^6}{Q^6}\ ,\ \ \ \ \
\delta a_1^{(6,\lambda)}\lesssim - 10^{-2}\,\, \frac{{\mathrm{GeV}}^6}{Q^6}\ .
\end{equation}
Despite the enhancement by one order of magnitude, the contribution of dimension-six operators (as well as the dimension-four operators) is negligible compared to the perturbative contribution,
Eq.~(\ref{pert}).
We conclude that we can safely neglect the contribution of the OPE to the
spectral functions, in comparison to perturbation theory.

\end{document}